\documentclass[aps,prb,reprint,superscriptaddress,longbibliography,nofootinbib]{revtex4-2}

\usepackage[utf8]{inputenc}
\usepackage{amsmath,amssymb,bm}
\usepackage{graphicx}
\usepackage{xcolor}
\usepackage{hyperref}
\hypersetup{
    colorlinks=true,
    linkcolor=blue,
    citecolor=blue,
    urlcolor=blue
}
\usepackage{physics}
\usepackage{mathtools}
\usepackage{mhchem}[version=4]
\usepackage{comment}
\usepackage{enumitem}

\begin{document}

\title{A Clockwork Quantum: Symmetry, Noise, and the Emergence of Quantum Order}

\author{Eric R. Bittner}
\email{bittner@uh.edu}
\affiliation{Department of Physics, University of Houston, Houston, TX 77204, USA}

\author{Bhavay Tyagi}
\affiliation{Department of Physics, University of Houston, Houston, TX 77204, USA}

\date{\today}

\begin{abstract}
We present a concise review and perspective on noise-induced synchronization and coherence protection in open quantum systems, with emphasis on  recent work involving coupled spins, oscillators, and anyons. When local environments exhibit internal correlations, the structure of the noise determines which collective modes become decoherence-protected. This leads to steady-state entanglement, phase locking, and exceptional points (EPs) in the Liouvillian spectrum, signaling a collapse of the mode basis and the emergence of non-dissipative stabilized dynamics.
Using a Lindblad framework, we show that symmetry in the noise correlations acts as a control parameter—protecting symmetric or antisymmetric modes depending on the sign of the correlation. In the pure-dephasing limit, coherence decay mirrors the Anderson–Kubo model, where the effective fluctuation strength scales as \( \sigma^2(1 \pm \xi) \), and the dynamical regime (Gaussian vs. Lorentzian) is set by the ratio \( \sigma / \gamma \). Thus, the environment not only drives decoherence but can also selectively suppress it through symmetry filtering.
We also revisit historical and conceptual origins of this idea, beginning with Huygens’ synchronized pendulum clocks and culminating in modern non-Hermitian dynamics. Correlated noise—though classically stochastic—can organize quantum dynamics and protect coherence without direct control over the system.
These insights offer a unifying view of synchronization in classical and quantum regimes, with implications for quantum sensing, engineered decoherence, and long-lived coherence in complex environments such as biological light-harvesting complexes or avian magnetoreception.
\end{abstract}

\maketitle

\section{Introduction}

Synchronization is a nearly universal feature of driven systems that share a common influence. From pendulum clocks and metronomes to power grids and coupled lasers, collective dynamics often emerge spontaneously in the presence of weak interactions or shared environments. In quantum systems, where coherence and dissipation compete on fundamental terms, the phenomenon of synchronization opens new avenues for probing decoherence, entanglement, and symmetry-breaking in open quantum dynamics.

\subsection{Historical Origins of Synchronization}

In 1665, Christiaan Huygens observed what he called “an odd kind of sympathy” between two pendulum clocks suspended from a common wooden beam. Over time, the clocks settled into a stable anti-phase oscillation—a phenomenon he initially attributed to air currents, but later recognized as a result of subtle mechanical coupling through the support. His correspondence from 1664–1665 documents what may be the first recorded instance of spontaneous synchronization between nonlinear oscillators~\cite{huygens1893correspondance}.

This early observation by Huygens inspired a series of investigations throughout the 18th and 19th centuries. John Ellicott reported detailed experiments on mutual influence between paired pendulum clocks in the Philosophical Transactions\cite{ellicott1739influence1,ellicott1739influence2}, demonstrating reproducible phase locking and frequency entrainment. Subsequent studies by Ellis and Korteweg\cite{ellis1873sympathetic,korteweg1906horloges} extended these findings, offering more refined interpretations of sympathetic resonance and mechanical coupling. The topic remained of interest into the early 20th century, with unsigned—but plausibly attributed—notes in Nature (1908, 1911)~\cite{korteweg1908synchronisation,korteweg1911synchronisation,korteweg_commentary} describing consistent observations of synchronized clock behavior.

\subsection{Modern Classical Perspectives}

The 20th century brought formal mathematical tools to the study of synchronization. Nonlinear dynamics, bifurcation theory, and coupled oscillator models revealed that phase locking and coherence arise naturally in weakly coupled dissipative systems. A particularly influential framework is the Kuramoto model, in which an ensemble of phase oscillators exhibits a phase transition from incoherence to global synchronization above a critical coupling strength~\cite{acebron2005kuramoto,strogatz2000kuramoto}. This model has become a paradigm for understanding emergent collective behavior in systems ranging from biological rhythms to superconducting circuits.

More recently, synchronization transitions have been reinterpreted through the lens of non-Hermitian and parity-time (\(\mathcal{PT}\)) symmetric physics. In this view, exceptional points in the system’s dynamical generator signal a symmetry-breaking transition, marking a change from stable to unstable collective modes. Comprehensive reviews by El‑Ganainy \emph{et al.}~\cite{elganainy2018nonhermitian} and Ashida \emph{et al.}~\cite{ashida2020nonhermitian}, along with foundational work by Bender~\cite{bender2007making}, have highlighted how non-Hermitian structures naturally arise in open, driven systems and lead to new forms of dynamical order. The work of Bennett \emph{et al.}~\cite{bennett2002huygens} on Huygens’s clocks offers a compelling bridge between these classical and modern views: their normal-mode analysis revealed that anti-phase locking arises from symmetry-selective coupling, a feature we find echoed in our quantum models as well.

Recent experimental work by Peña-Ramírez \emph{et al.} demonstrated Huygens-style synchronization using monumental pendulum clocks coupled through a flexible structure, providing modern validation of sympathetic resonance under noisy conditions~\cite{penaramires2016sympathy}.
Together, these contributions established a mechanistic basis for spontaneous synchronization and anticipated many of the features now recognized in nonlinear dynamical systems.

\subsection{Did nature think of this first?}
A growing body of experimental and theoretical research suggests that environmental structure—rather than always acting as a decohering influence—can play an active role in sustaining quantum coherence in complex biological systems. The pioneering 2D electronic spectroscopy study by Engel \emph{et al.} \cite{engel2007wavelike} first reported coherent oscillations in the Fenna–Matthews–Olson (FMO) complex, interpreted as evidence for wavelike excitonic energy transfer. Subsequent work by Panitchayangkoon \emph{et al.} \cite{panitchayangkoon2010long} extended this to physiological temperatures, raising the possibility that coherent energy transport might be biologically functional. Theoretical treatments by Ishizaki and Fleming \cite{ishizaki2009theoretical} used non-Markovian quantum master equations to show that structured, slow environmental modes can support long-lived coherence, while Chin \emph{et al.} \cite{chin2013role} demonstrated that non-equilibrium vibrational structures can enable electronic recoherence. Tiwari \emph{et al.} \cite{tiwari2013electronic} further argued that anticorrelated vibrational fluctuations across pigment sites contribute significantly to excitonic coupling and delocalization. These findings resonate strongly with the framework developed here: when the environmental correlations act through symmetrized noise channels, coherence can be selectively protected in specific normal modes. Our results place such protection on rigorous footing by connecting the correlation matrix to the irreducible representations of the system’s symmetry group, thereby providing a pathway for coherence stabilization through symmetry-preserving dissipation.

Recent work on avian navigation has explored how spin-correlated radical pairs in cryptochrome proteins may act as a quantum compass, with environmental correlations playing a crucial role in coherence lifetimes. The comprehensive review by Hore and Mouritsen \cite{hore2016radical} summarizes the biochemical and spin-dynamical foundations of the radical-pair mechanism. Rodgers and Hore \cite{rodgers2009chemical} first quantified how hyperfine interactions yield magnetic-field-dependent reaction yields. Efimova and Hore \cite{efimova2008role} analyzed the impact of electron–electron exchange and dipolar coupling on compass sensitivity. More recent theoretical advances by Player and Hore \cite{player2019viability} and Hiscock et al. \cite{hiscock2016quantum} emphasize how spatially correlated nuclear baths or radical scavengers can significantly enhance magnetic sensitivity by protecting coherence. These studies suggest that correlated environmental effects—notably those we model with symmetrized Lindblad channels—may indeed underlie the robust coherence observed in biological magnetoreception.

\subsection{Quantum Synchronization and Environmental Correlations}

At its core, synchronization reflects a flow of information through shared degrees of freedom. In Huygens’ experiment, the beam acted as a mechanical bus coupling the clocks. In open quantum systems, the environment plays an analogous role. When quantum subsystems couple to a common bath—especially one with internal correlations—the environment can induce phase locking, coherence, and even entanglement, despite acting solely through dissipative channels.

Central to our understanding of spectral diffusion and decoherence is the Anderson–Kubo stochastic model, introduced independently by Anderson \cite{anderson1954stochastic} and Kubo \cite{kubo1954resonance} in 1954. This foundational work describes how random frequency fluctuations—modeled as a Markovian jump process or continuous stochastic modulation—lead to characteristic broadening and lineshape changes in resonance spectra. By explicitly linking correlation times and fluctuation amplitudes to observable decoherence rates, the Anderson–Kubo model established the theoretical basis for interpreting noise-induced phenomena across magnetic resonance, optical spectroscopy, and single-molecule dynamics. This framework directly underpins our formulation of correlated dephasing via Ornstein–Uhlenbeck processes\cite{uhlenbeck1930brownian} and provides the analytical tools we leverage to connect noise correlations with mode-protected synchronization.

In recent work~\cite{tyagi2024noise,bittner2024correlated,bittner2025noise}, we formalized this intuition by modeling two qubits (or oscillators) locally coupled to environmental degrees of freedom. When the local noise sources were independent, the system exhibited rapid decoherence. But when the noise was drawn from a correlated Wiener process, we observed synchronization of the quantum phases, long-lived coherence, and the emergence of steady-state entanglement. These effects arise naturally in the Lindblad master equation when the dissipator includes off-diagonal terms that encode shared noise. The resulting generator exhibits a symmetry-breaking transition—analogous to \(\mathcal{PT}\)-symmetry breaking—marked by the appearance of exceptional points in the Liouvillian spectrum. In this regime, the long-time dynamics become dominated by a protected mode that can exhibit robust coherence or entanglement depending on the structure of the noise correlations.

We have also extended this framework to systems of anyons, where the exchange statistics interpolate continuously between bosons and fermions. In our recent study~\cite{bittner2025statistical}, we developed a Lindbladian formalism for coupled anyonic oscillators and demonstrated that the statistical phase \(\theta\) modifies the dissipative couplings in a nontrivial way. Depending on \(\theta\) and the structure of noise correlations, the system exhibits tunable coherence bifurcations, exceptional points, and symmetry-protected normal modes. These features appear directly in third-order spectroscopic observables and suggest experimental pathways to detecting fractional statistics using entirely classical probes.

Taken together, our studies point to a general principle: correlations within the environment act as a symmetry-breaking mechanism that selects dynamically protected subspaces. Within these subspaces, dissipation does not erase coherence—it stabilizes it. This principle may hold across a wide range of platforms, from spin qubits and coupled oscillators to topologically ordered matter, and suggests new avenues for exploiting environmental structure as a tool for quantum control.

\section{Theoretical Framework}

\subsection{Pure Dephasing Limit: Analytical Insight from the Spin–Boson Model}

To gain analytical traction and isolate the effects of correlated noise, we now consider a simplified version of the model in which each mode experiences only longitudinal coupling to its environment. 
The resulting Hamiltonian is:
\begin{equation}
    H(t) = \frac{\Delta}{2}(\sigma_z^{(1)} + \sigma_z^{(2)}) + \sigma_z^{(1)} E_1(t) + \sigma_z^{(2)} E_2(t),
    \label{eq:H_dephasing}
\end{equation}
which describes two independent spin-boson systems driven by correlated time-dependent noise fields \( E_1(t) \) and \( E_2(t) \). This setup mirrors the well-known Anderson–Kubo model for stochastic modulation of transition frequencies \cite{anderson1954stochastic,kubo1954resonance}, with the novel feature here being the cross-correlation between noise channels.

\subsection{Correlated Noise Structure}

We assume that the environmental fields \( \mathbf{E}_1(t) \) and \( \mathbf{E}_2(t) \) evolve according to a coupled Ornstein–Uhlenbeck (OU) process written in differential form:
\begin{equation}
    \dd{\mathbf{E}}(t) = -\gamma \mathbf{E}(t)\dd{t} + \mathbf{B} \cdot \dd{\mathbf{W}}(t),
\end{equation}
where \( \mathbf{E}(t) = \begin{pmatrix} E_1(t) \\ E_2(t) \end{pmatrix} \), and \( \dd{\mathbf{W}}(t) \) is a vector-valued Wiener process with correlation structure:
\begin{equation}
    \mathbb{E}\left[ \dd{\mathbf{W}}(t)\dd{\mathbf{W}}^\intercal(t') \right] = \boldsymbol{\Xi} \delta(t - t') \dd{t},
    \quad
    \boldsymbol{\Xi} =
    \begin{pmatrix}
    1 & \xi \\
    \xi & 1
    \end{pmatrix}.
\end{equation}
Here, the parameter \( \xi \in [-1, 1] \) controls the degree of instantaneous correlation between the two noise channels. The matrix \( \mathbf{B} \) determines the noise amplitude and, together with \( \boldsymbol{\Xi} \), sets the stationary variance of the process.
To analyze the resulting dynamics, we transform into a symmetry-adapted basis. Defining collective spin and noise coordinates:
\begin{align}
    \sigma_z^{(\pm)} &= \frac{1}{\sqrt{2}}(\sigma_z^{(1)} \pm \sigma_z^{(2)}), \\
    E_\pm(t) &= \frac{1}{\sqrt{2}}(E_1(t) \pm E_2(t)),
\end{align}
the interaction term becomes:
\begin{equation}
    H_{\text{int}}(t) = \sigma_z^{(+)} E_+(t) + \sigma_z^{(-)} E_-(t).
\end{equation}
Here, \( E_\pm(t) \) are independent OU processes with autocorrelation functions:
\begin{equation}
    \langle E_\pm(t) E_\pm(0) \rangle = \frac{\sigma^2}{2\gamma} (1 \pm \xi) e^{-\gamma t},
\end{equation}
showing that the symmetric and anti-symmetric combinations experience different effective fluctuation amplitudes, modulated by the correlation parameter \( \xi \).

We note that the product states \( \ket{00} \), \( \ket{01} \), \( \ket{10} \), and \( \ket{11} \) are eigenstates of the uncoupled system Hamiltonian. However, to analyze the effects of correlated dephasing noise, it is more natural to re-express these basis states in terms of linear combinations that are eigenstates of the collective spin operators \( \sigma_z^{(+)} = \sigma_z^{(1)} + \sigma_z^{(2)} \) and \( \sigma_z^{(-)} = \sigma_z^{(1)} - \sigma_z^{(2)} \). In particular, the superpositions
\begin{equation}
\ket{S} = \frac{1}{\sqrt{2}} ( \ket{01} + \ket{10} ), \quad
\ket{A} = \frac{1}{\sqrt{2}} ( \ket{01} - \ket{10} )
\end{equation}
form the symmetric and antisymmetric states in the one-excitation manifold. Dephasing noise enters through gap fluctuations that couple to the \( \sigma_z^{(i)} \) operators, and can be decomposed into fluctuations of the collective modes \( E_+(t) \) and \( E_-(t) \), corresponding to the symmetric and antisymmetric noise channels, respectively.

In this basis, the coherence \( \rho_{01,10} \) corresponds to an off-diagonal element in the \( \{ \ket{S}, \ket{A} \} \) subspace. Its dephasing rate is governed by the noise channel \( E_-(t) \), yielding
\begin{equation}
\Gamma_{01,10}(t) = \frac{\sigma^2}{\gamma^2}(1 - \xi)\left( \gamma t + e^{-\gamma t} - 1 \right),
\end{equation}
which vanishes in the limit \( \xi \to 1 \). Likewise, coherences between \( \ket{00} \) and \( \ket{11} \) couple to the symmetric noise channel \( E_+(t) \), and are governed by a rate proportional to \( (1 + \xi) \).

These results make explicit the symmetry-filtering role of noise correlations: in the fully anti-correlated case \( \xi = -1 \), the symmetric mode \( E_+ \) is protected from dephasing, while in the fully correlated case \( \xi = +1 \), it is the antisymmetric mode \( E_- \) that remains coherent. This confirms that noise correlations act as a symmetry filter—selectively protecting collective modes that are orthogonal to the dominant environmental coupling.

The behavior of coherence decay in our model exhibits strong parallels with the classic Anderson–Kubo theory of stochastic modulation, originally developed to explain homogeneous and inhomogeneous spectral broadening in nuclear magnetic resonance and optical spectroscopy. In that framework, a fluctuating transition frequency with characteristic strength \( \sigma \) and correlation time \( \tau_c = \gamma^{-1} \) leads to spectral line shapes that interpolate between Gaussian and Lorentzian depending on the ratio \( \sigma / \gamma \). In the slow modulation regime \( \gamma \ll \sigma \), the system samples many distinct configurations before dephasing completes, resulting in Gaussian lineshapes. In the fast modulation (motional narrowing) regime \( \gamma \gg \sigma \), the fluctuations average out, leading to a narrower Lorentzian profile.

In our case, the correlation coefficient \( \xi \) between the noise channels modifies the effective fluctuation amplitude experienced by each normal mode. Specifically, the symmetric mode \( E_+ = E_1 + E_2 \) samples fluctuations of strength \( \sigma_{\text{eff}}^2 \propto (1 + \xi) \), while the antisymmetric mode \( E_- = E_1 - E_2 \) samples fluctuations with \( \sigma_{\text{eff}}^2 \propto (1 - \xi) \). Thus, when \( \xi = 1 \), the antisymmetric mode becomes decoherence-free, entering a regime of motional narrowing with suppressed dephasing. Conversely, for \( \xi = -1 \), the symmetric mode becomes protected. Intermediate values of \( \xi \) interpolate between these extremes, determining which collective degrees of freedom exhibit dynamically protected coherence. Viewed through the lens of Anderson–Kubo theory, the correlation structure of environmental noise can shift the effective line-shape from Gaussian to Lorentzian via selective mode protection.

\subsection{Purity Decay of Bell States Under Correlated Dephasing}

We consider the four maximally entangled Bell states,
\begin{align}
\ket{\Phi^\pm} &= \tfrac{1}{\sqrt{2}}(\ket{00} \pm \ket{11}), \\
\ket{\Psi^\pm} &= \tfrac{1}{\sqrt{2}}(\ket{01} \pm \ket{10}),
\end{align}
expressed in the computational basis \(\{ \ket{00}, \ket{01}, \ket{10}, \ket{11} \}\), where \(\ket{ij} \equiv \ket{i}_1 \otimes \ket{j}_2\) denotes the joint state of two qubits. Each Bell state contains coherence between states that differ in total spin excitation number or spin localization, making them natural probes of how symmetry in the environmental noise affects decoherence.

Assuming that the system experiences only pure dephasing, the time-evolved density matrix retains its populations while the off-diagonal elements decay. Specifically, the relevant coherences decay according to:
\begin{align}
\ket{\Phi^\pm}: &\quad \rho_{00,11}(t) = \rho_{00,11}(0)\, e^{-\Gamma_{\Phi}(t)}, \\
\ket{\Psi^\pm}: &\quad \rho_{01,10}(t) = \rho_{01,10}(0)\, e^{-\Gamma_{\Psi}(t)},
\end{align}
with
\begin{equation}
\Gamma_{\Phi}(t) = 16\, \Gamma_+(t), \qquad
\Gamma_{\Psi}(t) = 16\, \Gamma_-(t),
\end{equation}
and
\begin{equation}
\Gamma_\pm(t) = \frac{\sigma^2}{\gamma^2}(1 \pm \xi)\left( \gamma t + e^{-\gamma t} - 1 \right).
\end{equation}

The purity, $\mathcal{P} = tr[\rho^2]$ at time \( t \) is then
\begin{equation}
\mathcal{P}_\Phi(t) = \frac{1}{2} \left( 1 + e^{-2\Gamma_\Phi(t)} \right), \qquad
\mathcal{P}_\Psi(t) = \frac{1}{2} \left( 1 + e^{-2\Gamma_\Psi(t)} \right).
\end{equation}

These expressions show explicitly how the purity of each Bell state depends on the sign and magnitude of the correlation parameter \( \xi \). In the limit \( \xi \to +1 \), \( \Gamma_+ \) is maximized and \( \Gamma_- \to 0 \); conversely, for \( \xi \to -1 \), the roles are reversed.

This selectivity can be understood heuristically by examining how each Bell state encodes coherence across different energy levels. The \(\Phi^\pm\) states involve coherence between the doubly excited state \(\ket{11}\) and the ground state \(\ket{00}\), and are thus sensitive to fluctuations in the \emph{total energy gap}, which is proportional to \(E_1(t) + E_2(t)\). As a result, their dephasing is governed by the symmetric noise channel \(E_+(t)\), and these states decohere most rapidly under fully correlated noise (\(\xi = +1\)). Conversely, the \(\Psi^\pm\) states involve coherence between the singly excited configurations \(\ket{10}\) and \(\ket{01}\), and are therefore sensitive to fluctuations in the \emph{energy difference} between the two sites, i.e., \(E_1(t) - E_2(t)\). These states decohere via the antisymmetric noise channel \(E_-(t)\), and are most susceptible to dephasing under fully anti-correlated noise (\(\xi = -1\)). This symmetry-selective decay mechanism, evident here in a pure dephasing model, echoes our earlier results obtained using a full Redfield treatment with structured spectral densities.

While the coherent dynamics have been switched off here for clarity, this model illustrates a key point: dissipation is not uniformly distributed across the Hilbert space. Instead, the structure of the environment—and in particular, the symmetry of the noise correlation matrix—selects which superpositions are robust and which are fragile. This lays a foundation for more general results in synchronization theory, and suggests possible applications in noise-protected quantum protocols.


\subsection{Correlated Lindblad equations}
While the stochastic description provides physical intuition and analytical tractability in certain limits, it is often more convenient to work directly within the Lindblad formalism when modeling Markovian open quantum systems. The Lindblad master equation assumes weak system–bath coupling, separation of timescales (i.e., fast bath decorrelation relative to system dynamics), and complete positivity of the evolution map. Under these assumptions, the system’s density matrix \( \rho(t) \) evolves according to:
\begin{equation}
    \dv{\rho}{t} = -\frac{i}{\hbar}[H_0, \rho] + \sum_\alpha \left( L_\alpha \rho L_\alpha^\dagger - \frac{1}{2} \{ L_\alpha^\dagger L_\alpha, \rho \} \right),
    \label{eq:lindblad}
\end{equation}
where \( \{L_\alpha\} \) are a set of Lindblad (or jump) operators that describe system–environment coupling pathways. These operators encode both the spectral properties of the bath and the symmetry of the coupling, and may act locally or collectively on multiple degrees of freedom.

In typical open-system models, each subsystem is coupled to an independent bath, and the Lindblad operators act locally:
\begin{equation}
    L_1 = \sqrt{\gamma_1} \, A_1, \quad L_2 = \sqrt{\gamma_2} \, A_2,
\end{equation}
where \( A_i \) is a Hermitian or non-Hermitian system operator (e.g., \( \sigma_z^{(i)} \), \( \sigma_-^{(i)} \), etc.) and \( \gamma_i \) sets the corresponding relaxation rate. However, this picture excludes a key ingredient of many realistic environments: correlations between the baths.

To incorporate bath correlations into the Lindblad framework, we extend the dissipation model by allowing the noise acting on different subsystems to be statistically correlated. As shown previously, these correlations can be described by a noise covariance matrix \( \boldsymbol{\Xi} \). To embed this structure into the Lindblad formalism, we introduce a set of \emph{correlated Lindblad operators} defined via linear combinations of local system operators:
\begin{equation}
    \mathbf{L} =
    \begin{pmatrix}
        L_1 \\
        L_2
    \end{pmatrix}
    =
    \mathbf{A} \cdot \mathbf{C},
    \label{eq:correlated_L}
\end{equation}
where \( \mathbf{A} = \begin{pmatrix} A_1 & A_2 \end{pmatrix}^\intercal \) contains the local system operators, and \( \mathbf{C} \) is a noise-correlation matrix (e.g., Cholesky decomposition of \( \boldsymbol{\Xi} \)) satisfying:
\begin{equation}
    \boldsymbol{\Xi} = \mathbf{C} \mathbf{C}^\dagger.
\end{equation}

The resulting dissipator becomes:
\begin{equation}
    \mathcal{D}[\rho] = \sum_{i,j} \Xi_{ij} \left( A_i \rho A_j^\dagger - \frac{1}{2} \{ A_j^\dagger A_i, \rho \} \right),
    \label{eq:correlated_dissipator}
\end{equation}
which generalizes the standard form by including off-diagonal contributions that couple subsystems \( i \) and \( j \) via correlated fluctuations. This formulation is manifestly trace-preserving and completely positive as long as \( \boldsymbol{\Xi} \) is positive semidefinite.

In the context of our two-mode system, the relevant system operators are taken to be:
\begin{equation}
    A_1 = \sigma_z^{(1)}, \quad A_2 = \sigma_z^{(2)},
\end{equation}
so that the dissipator models longitudinal dephasing noise with tunable correlation. When \( \Xi_{12} = 0 \), the noise acts independently on each qubit or oscillator; when \( \Xi_{12} \neq 0 \), the dissipation acts jointly on both modes, inducing nonlocal effects such as synchronized decay and symmetry-protected subspaces.

This formalism connects smoothly to the stochastic differential equation picture introduced earlier. In fact, the same covariance matrix \( \boldsymbol{\Xi} \) that defines the cross-correlation of the Wiener processes in the Ito equations now determines the strength and structure of the Lindblad dissipator. This correspondence allows us to move fluidly between analytical stochastic treatments and full quantum dynamical simulations, depending on the specific regime and observables of interest.





\subsection{Exceptional Points  in the Non-Resonant Oscillator Model}

To illustrate the spectral implications of correlated dissipation, we consider a paradigmatic open quantum system comprising two linearly coupled harmonic oscillators with unequal frequencies. The Hamiltonian is given by
\begin{equation}
H = \omega_1 a_1^\dagger a_1 + \omega_2 a_2^\dagger a_2 + J (a_1^\dagger a_2 + a_2^\dagger a_1),
\end{equation}
where \( \omega_1 \neq \omega_2 \) accounts for detuning between the two modes, and \( J \in \mathbb{R} \) denotes the coherent exchange coupling. Each oscillator is coupled to a finite-temperature environment through amplitude damping operators,
\begin{equation}
L_1 = \sqrt{\gamma (n_1+1)}\, a_1, \qquad L_2 = \sqrt{\gamma (n_2+1)}\, a_2,
\end{equation}
and amplitude creation operators,
\begin{equation}
L_3 = \sqrt{\gamma n_1}\, a_1^\dagger, \qquad L_4 = \sqrt{\gamma n_2}\, a_2^\dagger,
\end{equation}
where $n_i$ sets the thermal occupation of each mode.

\begin{figure}
    \centering
    \includegraphics[width=\linewidth]{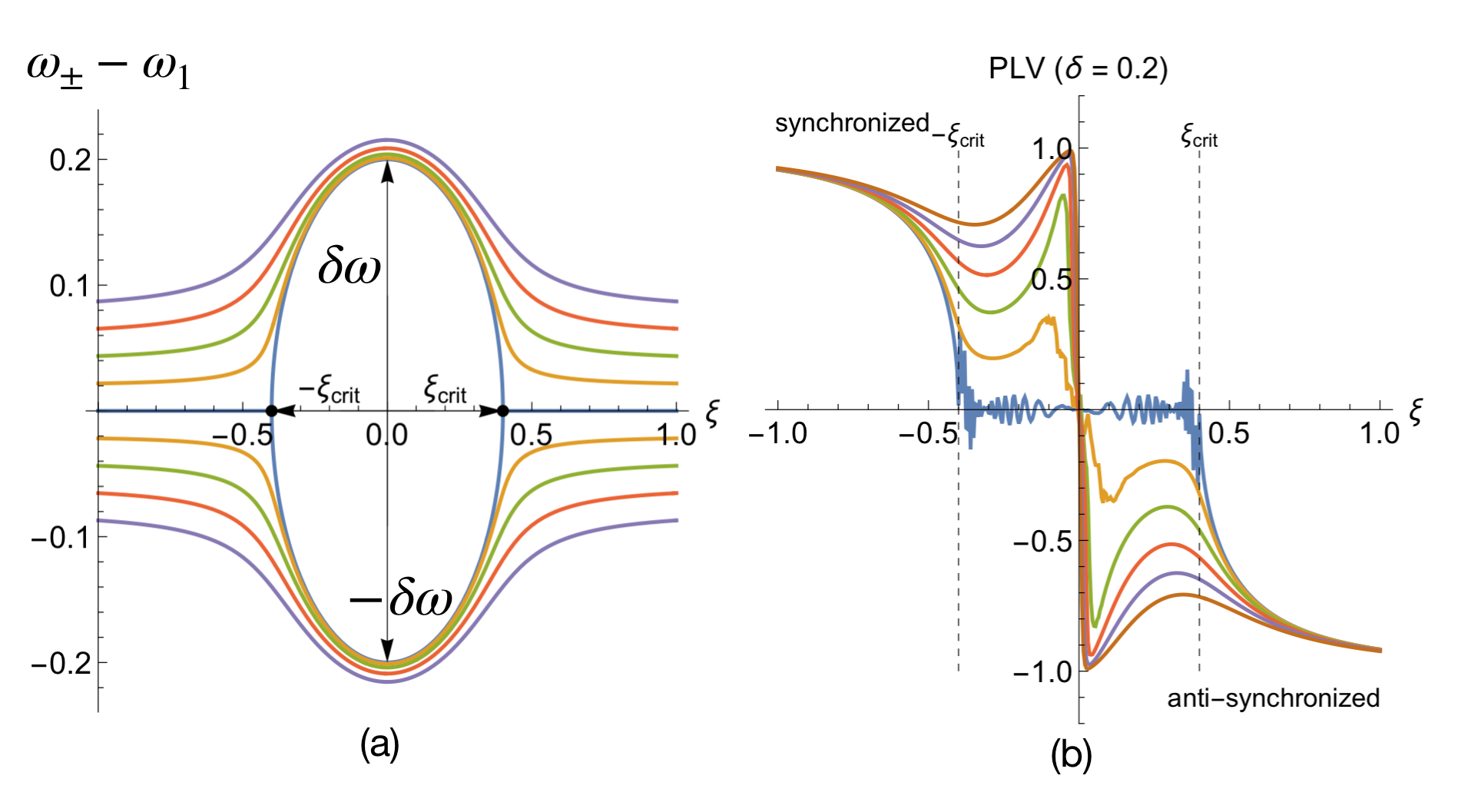}
  \caption{
    \textbf{Dynamical signatures of environmentally induced synchronization in two coupled oscillators under correlated amplitude damping.}
    (a) Imaginary parts of the eigenvalues \( \lambda_\pm \) of the drift matrix \( \mathbf{W} \), showing the mode frequencies as a function of the noise correlation parameter \( \xi \). At fixed detuning \( \Delta\omega \), the eigenfrequencies converge at a critical correlation threshold determined by \( J \) and \( \gamma \), signaling the emergence of synchronized phase evolution. 
    (b) Time-averaged phase-locking order parameter \( \langle \cos[\phi(t)] \rangle \), computed by averaging over an ensemble of initial conditions. As \( \xi \to \pm1 \), the relative phase stabilizes, confirming the environment-driven transition to coherence.
    This figure illustrates how symmetry in the environmental correlation selectively protects collective modes and leads to robust phase alignment even in the absence of direct interaction.
    }
    \label{fig:2}
\end{figure}

The first-moment dynamics of the system, in the Heisenberg picture, obey a linear differential equation of the form
\begin{equation}
    \dv{}{t} \langle \mathbf{v}(t) \rangle = \mathbf{W} \langle \mathbf{v}(t) \rangle,
\end{equation}
where \( \mathbf{v}(t) \) is the column vector of system observables (e.g., Bloch vector components or bosonic mode amplitudes), and \( \mathbf{W} \) is the non-Hermitian drift matrix defined by the coherent Hamiltonian contribution and dissipative Lindblad terms. The formal solution is
\begin{equation}
    \langle \mathbf{v}(t) \rangle = e^{\mathbf{W}t} \langle \mathbf{v}(0) \rangle,
\end{equation}
where the dynamics are governed by the eigenvalues \( \lambda_\pm \) of \( \mathbf{W} \), 
\begin{equation}
    \lambda_{\pm} = -i\bar{\omega} - \gamma \pm \sqrt{ \left( \frac{-i \Delta \omega}{2} \right)^2 - (J + i \gamma \xi)^2 }.
\end{equation}
Here \( \bar{\omega} = (\omega_1 + \omega_2)/2 \) and \( \Delta \omega = \omega_1 - \omega_2 \). The real parts of \( \lambda_\pm \) determine the decay rates, while the imaginary parts correspond to the oscillation frequencies of the dynamical modes.

To visualize how environmental correlations impact the coherent structure, we compute the imaginary parts of \( \lambda_\pm \) as a function of the correlation parameter \( \xi \), shown in Fig.~\ref{fig:2}(a)
for increasing values of 
intermode coupling $J$.
As \( \xi \) increases, the two eigenfrequencies merge at a critical coupling, resulting in degenerate phase evolution and an emergent synchronized mode. This merging behavior underlies the transition to phase locking.

To quantify this effect, we compute the time-averaged phase difference between the two modes across a statistical ensemble of initial conditions. Specifically, defining the relative phase \( \phi(t) = \arg[a_1(t)] - \arg[a_2(t)] \), we compute the ensemble average
\begin{equation}
    \langle \cos[\phi(t)] \rangle = \mathrm{Re} \left\langle \frac{a_1^*(t)a_2(t)}{|a_1(t)||a_2(t)|} \right\rangle,
\end{equation}
as a function of \( \xi \). The resulting behavior (in Fig.~\ref{fig:2}(b)) shows a sharp onset of phase correlation as \( \xi \to \pm 1 \), confirming that synchronization is driven by environmental symmetry structure even in the absence of direct coupling.





Exceptional points (EPs) arise when the eigenvalues of \( \mathbf{W} \) coalesce, which occurs when the discriminant of the characteristic polynomial vanishes. This condition is satisfied when
\[
\left(2\gamma \xi + 2iJ\right)^2 = \Delta\omega^2.
\] 
When \( J = 0 \), the condition reduces to \( \xi_{\mathrm{EP}} = \pm \Delta\omega / (2\gamma) \), and the exceptional points lie on the real axis. However, for nonzero coupling \( J \neq 0 \), the EPs move into the complex-$\xi$ plane, and the condition becomes
\[
\xi_{\mathrm{EP}} = \frac{1}{2\gamma} \left( \pm \Delta\omega - 2iJ \right)
\]
This defines a pair of isolated branch points in the complex-\( \xi \) plane located at symmetric values of \( \operatorname{Re}(\xi) = \pm \Delta\omega / (2\gamma) \), and vertically offset along the imaginary axis by \( \operatorname{Im}(\xi) = -J/\gamma \). These complex EPs mark the onset of dynamical non-diagonalizability and the collapse of the mode basis, leading to distinct spectral and phase signatures in the evolution of coherence and population.

This exceptional seam encodes the topology of the non-Hermitian parameter space. Encircling it in the complex plane causes the eigenvalues to undergo exchange, acquiring a nontrivial geometric phase. The real-valued correlation parameter \( \xi \) thus plays the role of a hidden symmetry axis, and the complex continuation reveals a dynamical analog of a phase transition—where symmetry-protected modes coalesce and bifurcate, marking a boundary between coherent and incoherent dynamics.
This defines a dissipation-driven symmetry-breaking threshold, below which the system remains in the \(\mathcal{PT}\)-symmetric (oscillatory) regime and above which the modes hybridize and decay asymmetrically. The resulting exceptional point reflects a competition between coherent detuning and noise-induced dissipation.

\subsection{Quantum versus Classical Correlation}

An important conceptual distinction in our model is the difference between classical correlations—those induced by shared noise—and genuinely quantum correlations, such as entanglement or quantum discord.  We now ask a more foundational question: \emph{how much quantum information is actually shared between the components?} Intuitively, one might expect no quantum correlation between subsystems unless a coherent exchange term (\( J \)) is present. However, in both our previous studies of qubit dimers~\cite{bittner2024correlated,tyagi2024noise} and bosonic oscillators~\cite{bittner2025noise}, we found that correlations within the bath—encoded by the parameter \(\xi\)—can induce nonclassical correlations even when \( J = 0 \). This arises because the system becomes effectively entangled with a common environmental mode that is itself correlated across sites.

To quantify the genuinely quantum component of shared information, we introduced a measure \(\mathcal{Q}\), defined as the difference between the total mutual information \( I(A:B) \) and the maximal classical information that remains after eliminating quantum coherences between \( A \) and \( B \). In our Gaussian oscillator model, this quantity can be rigorously computed using second R\'enyi entropies, exploiting the fact that these obey strong subadditivity for Gaussian states. The R\'enyi-2 mutual information is expressed as
\[
I_2(A:B) = \frac{1}{2} \ln\left( \frac{\det \Theta_A \det \Theta_B}{\det \Theta_{AB}} \right),
\]
where \(\Theta_{AB}\) is the joint covariance matrix and \(\Theta_{A,B}\) are the local marginals.

To define a classical reference, we suppress the off-diagonal blocks (\(\Gamma\)) linking \(A\) and \(B\), and recompute the mutual information. The difference defines a lower bound to the shared quantum information:
\[
\mathcal{Q} = I_2(A:B) - I_2^{\text{class}}(A:B).
\]
This approach ensures that \(\mathcal{Q}\) vanishes if and only if the full covariance matrix is block-diagonal—i.e., when no quantum coherence or discord is present.

Remarkably, our calculations show that \(\mathcal{Q}\) increases significantly as \(|\xi| \to 1\), even when \(J = 0\). In this limit, the environmental correlation acts as a nonlocal quantum channel, enforcing phase alignment and generating nonzero mutual information that is purely quantum in origin.

Taken together, our definition of the quantumness, \( \mathcal{Q} \), serves as a rigorous, computable lower bound on shared quantum information in noise-driven systems. They also reinforce the broader theme of this work: that entanglement and coherence can arise from symmetry and structure in the environment, rather than from direct Hamiltonian coupling.

\begin{figure}[t]
    \centering
    \includegraphics[width=0.45\textwidth]{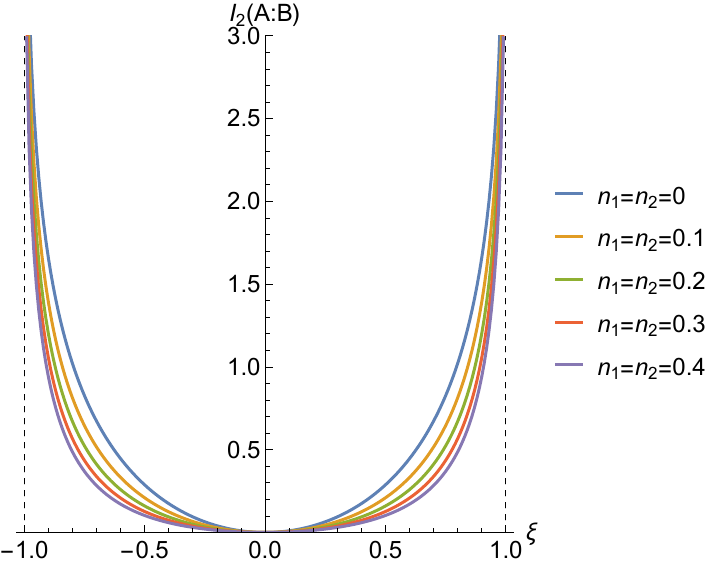}
    \caption{Illustration of \(\mathcal{Q}\), the lower bound on quantum correlation, defined as the difference between total and classical R\'enyi-2 mutual information. Figure reproduced from Ref.~\citenum{bittner2025noise}.}
    \label{fig:Q_discord}
\end{figure}

\section{Discussion}

The central finding of this work is that correlated noise—often viewed as a detrimental or incoherent influence—can, in fact, act as a symmetry-selective mechanism that shapes the dissipation landscape of a quantum system. In particular, we have shown that synchronization emerges not from direct coupling between subsystems, but rather from the structure of the noise correlation itself. This effect is robust and manifests across various physical realizations, including qubit pairs, coupled oscillators, and anyonic systems with statistical deformation.

In the two-mode system, the effect is clearest. When two qubits or oscillators are subject to fully \emph{anti-correlated} noise, the \emph{symmetric} normal mode becomes dynamically decoupled from the environment, experiencing significantly reduced dissipation. Conversely, under fully \emph{correlated} noise, the \emph{anti-symmetric} mode is protected. This selection of dissipation-protected subspaces is not arbitrary; it reflects a symmetry constraint imposed by the cross terms in the Lindblad dissipator, which preserve certain irreducible representations of the system's underlying symmetry group.

An important insight from the Anderson–Kubo perspective is that the spectral character of decoherence—whether it appears as inhomogeneous (Gaussian) or homogeneous (Lorentzian)—is governed by the interplay between the fluctuation strength \( \sigma \) and correlation time \( \tau_c = \gamma^{-1} \). In our model, the correlation parameter \( \xi \) modifies this balance in a symmetry-resolved way. The effective fluctuation strength for each normal mode becomes \( \sigma^2_{\text{eff}} \propto \sigma^2 (1 \pm \xi) \), thereby shifting the system across the Gaussian–Lorentzian boundary depending on the noise symmetry.

Specifically, a mode becomes \textit{effectively motionally narrowed} when \( \gamma \gg \sigma \sqrt{1 \pm \xi} \), suppressing dephasing and yielding Lorentzian-like lineshapes. In contrast, when \( \gamma \ll \sigma \sqrt{1 \pm \xi} \), the fluctuations are effectively static during the coherence window, resulting in Gaussian-like broadening. Thus, by tuning \( \xi \), one can manipulate the decoherence regime for specific symmetry sectors—enhancing or suppressing coherence decay without altering the system Hamiltonian. This symmetry-resolved tunability may provide a powerful handle for spectral control in cavity QED systems, excitonic complexes, or engineered quantum devices with structured noise.

This mechanism generalizes to larger networks of coupled quantum systems, where the environmental correlations can be described by a \emph{noise correlation matrix}, or more abstractly, as an adjacency matrix on a graph. When the noise structure is modeled by a cycle graph $C_N$, the system inherits the symmetry group of the graph—namely, the cyclic group $\mathbb{Z}_N$. The collective modes of the system then form a basis of irreducible representations (irreps) of this group. Each irrep corresponds to a collective excitation, and its relaxation rate is determined by how it overlaps with the symmetry of the noise correlation pattern.

Our simulations and analytic results suggest a general principle: \emph{the irrep associated with the smallest eigenvalue of the noise correlation graph exhibits the weakest dissipation and dominates the long-time dynamics}. In this sense, the environment acts as a filter, projecting the system's evolution onto the slowest-damped subspace. This subspace often corresponds to a synchronized or entangled collective mode, even in the absence of direct interaction between individual subsystems.

This framework provides a natural path toward scalable synchronization in quantum systems. By engineering the correlation structure of the environment—e.g., via shared phonon baths, common photonic modes, or structured reservoirs—it becomes possible to design networks with tailored dissipation hierarchies. In practice, this could be used to stabilize specific quantum states, enhance metrological sensitivity, or protect entangled modes in quantum communication protocols.

The structure of our model shares important conceptual parallels with the classical Kuramoto model of coupled oscillators~\cite{acebron2005kuramoto, ott2008low, lee2011quantum}. In the Kuramoto framework, a population of oscillators with distributed natural frequencies synchronizes above a critical coupling threshold due to sinusoidal phase interactions. The resulting dynamics are characterized by a transition from incoherent to phase-locked behavior, with a collective order parameter quantifying the degree of synchronization.

In our quantum setting, synchronization and coherence protection arise not from sinusoidal coupling but from the interplay between coherent exchange terms and environmental noise correlations. The correlation parameter \( \xi \) plays a role analogous to coupling strength in Kuramoto models, selecting symmetry-protected subspaces and enabling long-lived phase correlations between subsystems. The emergence of exceptional points (EPs) and the mode selectivity induced by correlated noise echo the critical transition observed in the Kuramoto model, where a macroscopic fraction of oscillators phase-lock due to shared dynamical influence.

Furthermore, just as generalized Kuramoto models with common noise exhibit enhanced synchronization~\cite{teramae2004robustness, nakao2007noise}, our model demonstrates that structured environmental fluctuations can induce coherence and entanglement even in the absence of direct coupling. This unifies classical and quantum views of synchronization under a broader framework in which shared stochasticity acts as the organizing principle.

We also note that our findings resonate with a broad body of literature showing that a shared environment alone can generate quantum correlations without direct system–system coupling. In seminal work, Braun demonstrated that two non-interacting qubits immersed in a common thermal bath can become transiently entangled~\cite{braun2002creation}, and Benatti et al.\ showed similar effects in the steady-state regime of Gaussian oscillator modes ~\cite{benatti2006entangling}. More recent studies have explored how classical, structured noise can generate and preserve quantum discord even in the absence of coherent coupling~\cite{benedetti2012effects}. Our analysis extends and refines these results in several ways:
\begin{itemize}
\item   We introduce a \emph{controllable noise-correlation parameter} \(\xi\), allowing one to tune between symmetric and anti-symmetric decoherence-free subspaces and directly observe phase-locking and coherence protection.

\item  We quantify non-classicality via a R\'enyi-2 mutual information framework, yielding a computable lower bound \(\mathcal{Q}\) on quantum discord even for \(J = 0\).

\item  We reveal that in the limit \(|\xi| \to 1\), quantum correlations not only appear but can persist indefinitely in the absence of Hamiltonian coupling, demonstrating that dissipation itself, when structured, can act as a conduit for quantum information.
\end{itemize}

While the primary focus of this paper is on spin and bosonic systems, we have recently extended this framework to explore synchronization phenomena in systems obeying fractional statistics~\cite{bittner2025statistical}. In that work, we developed a Lindblad formalism for anyonic oscillators, in which the commutation relations between creation and annihilation operators acquire a phase determined by the statistical angle \( \theta \). This deformation of the algebra has profound consequences for the structure of the Liouvillian and the nature of exceptional points (EPs). Specifically, we find that the locations of EPs in the complex plane—and the symmetry of the protected modes—are strongly dependent on \( \theta \), leading to tunable bifurcations in coherence and population dynamics. 

The presence of fractional statistics enriches the synchronization landscape by introducing interference effects that are neither purely bosonic nor fermionic in nature. As the noise correlations \( \xi \) are varied, the system explores distinct decoherence-free subspaces whose structure depends sensitively on \( \theta \). These features leave observable signatures in third-order nonlinear spectroscopic signals, including phase-dependent enhancements in two-dimensional Fourier spectra and shifts in rephasing vs.\ non-rephasing pathways. As such, nonlinear spectroscopy provides a potential probe of anyonic coherence dynamics without requiring topological braiding operations or strong-coupling measurement schemes. 

This extension underscores the generality of our approach: the symmetry of the environmental correlations, together with the statistical structure of the system, determines which dynamical modes are protected and how coherence is distributed across the network. We expect that further exploration of noise-induced order in anyonic and topologically nontrivial systems will reveal new mechanisms for engineering quantum correlations in noisy environments.

Finally, it is worth noting that while the noise itself may be classical (e.g., modeled as Ornstein–Uhlenbeck processes), the induced correlations and synchronization are distinctly quantum. The long-lived coherence, quantum discord, and mutual information observed in the steady state reflect a subtle interplay between classical correlations in the bath and quantum correlations in the system. This resonates with foundational questions about the role of hidden variables—here realized not as intrinsic features of the wavefunction, but as structural features of the environment that select and stabilize emergent quantum order.

\begin{acknowledgments}
This work was supported by the National Science Foundation under CHE-2404788 and the Robert A. Welch Foundation (E-1337).
\end{acknowledgments}

\bibliographystyle{apsrev4-2}
\bibliography{References_local}

\begin{thebibliography}{39}%
\makeatletter
\providecommand \@ifxundefined [1]{%
 \@ifx{#1\undefined}
}%
\providecommand \@ifnum [1]{%
 \ifnum #1\expandafter \@firstoftwo
 \else \expandafter \@secondoftwo
 \fi
}%
\providecommand \@ifx [1]{%
 \ifx #1\expandafter \@firstoftwo
 \else \expandafter \@secondoftwo
 \fi
}%
\providecommand \natexlab [1]{#1}%
\providecommand \enquote  [1]{``#1''}%
\providecommand \bibnamefont  [1]{#1}%
\providecommand \bibfnamefont [1]{#1}%
\providecommand \citenamefont [1]{#1}%
\providecommand \href@noop [0]{\@secondoftwo}%
\providecommand \href [0]{\begingroup \@sanitize@url \@href}%
\providecommand \@href[1]{\@@startlink{#1}\@@href}%
\providecommand \@@href[1]{\endgroup#1\@@endlink}%
\providecommand \@sanitize@url [0]{\catcode `\\12\catcode `\$12\catcode `\&12\catcode `\#12\catcode `\^12\catcode `\_12\catcode `\%12\relax}%
\providecommand \@@startlink[1]{}%
\providecommand \@@endlink[0]{}%
\providecommand \url  [0]{\begingroup\@sanitize@url \@url }%
\providecommand \@url [1]{\endgroup\@href {#1}{\urlprefix }}%
\providecommand \urlprefix  [0]{URL }%
\providecommand \Eprint [0]{\href }%
\providecommand \doibase [0]{https://doi.org/}%
\providecommand \selectlanguage [0]{\@gobble}%
\providecommand \bibinfo  [0]{\@secondoftwo}%
\providecommand \bibfield  [0]{\@secondoftwo}%
\providecommand \translation [1]{[#1]}%
\providecommand \BibitemOpen [0]{}%
\providecommand \bibitemStop [0]{}%
\providecommand \bibitemNoStop [0]{.\EOS\space}%
\providecommand \EOS [0]{\spacefactor3000\relax}%
\providecommand \BibitemShut  [1]{\csname bibitem#1\endcsname}%
\let\auto@bib@innerbib\@empty
\bibitem [{\citenamefont {Nijhoff}(1893)}]{huygens1893correspondance}%
  \BibitemOpen
  \bibinfo {editor} {\bibfnamefont {M.}~\bibnamefont {Nijhoff}},\ ed.,\ \href@noop {} {\emph {\bibinfo {title} {Oeuvres Compl{\`e}tes de Christiaan Huygens, Tome V: Correspondance 1664--1665}}}\ (\bibinfo  {publisher} {Martinus Nijhoff / La Soci{\'e}t{\'e} Hollandaise des Sciences},\ \bibinfo {address} {The Hague},\ \bibinfo {year} {1893})\BibitemShut {NoStop}%
\bibitem [{\citenamefont {Ellicott}(1739{\natexlab{a}})}]{ellicott1739influence1}%
  \BibitemOpen
  \bibfield  {author} {\bibinfo {author} {\bibfnamefont {J.}~\bibnamefont {Ellicott}},\ }\href@noop {} {\bibfield  {journal} {\bibinfo  {journal} {Philosophical Transactions}\ }\textbf {\bibinfo {volume} {41}},\ \bibinfo {pages} {126} (\bibinfo {year} {1739}{\natexlab{a}})}\BibitemShut {NoStop}%
\bibitem [{\citenamefont {Ellicott}(1739{\natexlab{b}})}]{ellicott1739influence2}%
  \BibitemOpen
  \bibfield  {author} {\bibinfo {author} {\bibfnamefont {J.}~\bibnamefont {Ellicott}},\ }\href@noop {} {\bibfield  {journal} {\bibinfo  {journal} {Philosophical Transactions}\ }\textbf {\bibinfo {volume} {41}},\ \bibinfo {pages} {128} (\bibinfo {year} {1739}{\natexlab{b}})}\BibitemShut {NoStop}%
\bibitem [{\citenamefont {Ellis}(1873)}]{ellis1873sympathetic}%
  \BibitemOpen
  \bibfield  {author} {\bibinfo {author} {\bibfnamefont {W.}~\bibnamefont {Ellis}},\ }\href@noop {} {\bibfield  {journal} {\bibinfo  {journal} {Monthly Notices of the Royal Astronomical Society}\ }\textbf {\bibinfo {volume} {33}},\ \bibinfo {pages} {480} (\bibinfo {year} {1873})}\BibitemShut {NoStop}%
\bibitem [{\citenamefont {Korteweg}(1906)}]{korteweg1906horloges}%
  \BibitemOpen
  \bibfield  {author} {\bibinfo {author} {\bibfnamefont {D.~J.}\ \bibnamefont {Korteweg}},\ }in\ \href@noop {} {\emph {\bibinfo {booktitle} {Archives N{\'e}erlandaises des Sciences Exactes et Naturelles, Tome XI}}}\ (\bibinfo  {publisher} {La Soci{\'e}t{\'e} Hollandaise des Sciences {\`a} Harlem, The Hague},\ \bibinfo {year} {1906})\ pp.\ \bibinfo {pages} {273--295}\BibitemShut {NoStop}%
\bibitem [{\citenamefont {Korteweg}(1908)}]{korteweg1908synchronisation}%
  \BibitemOpen
  \bibfield  {author} {\bibinfo {author} {\bibfnamefont {D.~J.}\ \bibnamefont {Korteweg}},\ }\href {https://doi.org/10.1038/078353a0} {\bibfield  {journal} {\bibinfo  {journal} {Nature}\ }\textbf {\bibinfo {volume} {78}},\ \bibinfo {pages} {353} (\bibinfo {year} {1908})}\BibitemShut {NoStop}%
\bibitem [{\citenamefont {Korteweg}(1911)}]{korteweg1911synchronisation}%
  \BibitemOpen
  \bibfield  {author} {\bibinfo {author} {\bibfnamefont {D.~J.}\ \bibnamefont {Korteweg}},\ }\href {https://doi.org/10.1038/085516a0} {\bibfield  {journal} {\bibinfo  {journal} {Nature}\ }\textbf {\bibinfo {volume} {85}},\ \bibinfo {pages} {516} (\bibinfo {year} {1911})}\BibitemShut {NoStop}%
\bibitem [{kor()}]{korteweg_commentary}%
  \BibitemOpen
  \href@noop {} {}\bibinfo {note} {Note: {A}lthough unsigned, the 1908 and 1911 \textit{Nature} articles cited above on clock synchronization are attributed to Korteweg in modern citation metadata.}\BibitemShut {Stop}%
\bibitem [{\citenamefont {Acebr{\'o}n}\ \emph {et~al.}(2005)\citenamefont {Acebr{\'o}n}, \citenamefont {Bonilla}, \citenamefont {Vicente}, \citenamefont {Ritort},\ and\ \citenamefont {Spigler}}]{acebron2005kuramoto}%
  \BibitemOpen
  \bibfield  {author} {\bibinfo {author} {\bibfnamefont {J.~A.}\ \bibnamefont {Acebr{\'o}n}}, \bibinfo {author} {\bibfnamefont {L.~L.}\ \bibnamefont {Bonilla}}, \bibinfo {author} {\bibfnamefont {C.~J.~P.}\ \bibnamefont {Vicente}}, \bibinfo {author} {\bibfnamefont {F.}~\bibnamefont {Ritort}},\ and\ \bibinfo {author} {\bibfnamefont {R.}~\bibnamefont {Spigler}},\ }\href@noop {} {\bibfield  {journal} {\bibinfo  {journal} {Reviews of Modern Physics}\ }\textbf {\bibinfo {volume} {77}},\ \bibinfo {pages} {137} (\bibinfo {year} {2005})}\BibitemShut {NoStop}%
\bibitem [{\citenamefont {Strogatz}(2000)}]{strogatz2000kuramoto}%
  \BibitemOpen
  \bibfield  {author} {\bibinfo {author} {\bibfnamefont {S.~H.}\ \bibnamefont {Strogatz}},\ }\href {https://doi.org/10.1016/S0167-2789(00)00094-4} {\bibfield  {journal} {\bibinfo  {journal} {Physica D: Nonlinear Phenomena}\ }\textbf {\bibinfo {volume} {143}},\ \bibinfo {pages} {1} (\bibinfo {year} {2000})}\BibitemShut {NoStop}%
\bibitem [{\citenamefont {El-Ganainy}\ \emph {et~al.}(2018)\citenamefont {El-Ganainy}, \citenamefont {Makris}, \citenamefont {Khajavikhan}, \citenamefont {Musslimani}, \citenamefont {Rotter},\ and\ \citenamefont {Christodoulides}}]{elganainy2018nonhermitian}%
  \BibitemOpen
  \bibfield  {author} {\bibinfo {author} {\bibfnamefont {R.}~\bibnamefont {El-Ganainy}}, \bibinfo {author} {\bibfnamefont {K.~G.}\ \bibnamefont {Makris}}, \bibinfo {author} {\bibfnamefont {M.}~\bibnamefont {Khajavikhan}}, \bibinfo {author} {\bibfnamefont {Z.~H.}\ \bibnamefont {Musslimani}}, \bibinfo {author} {\bibfnamefont {S.}~\bibnamefont {Rotter}},\ and\ \bibinfo {author} {\bibfnamefont {D.~N.}\ \bibnamefont {Christodoulides}},\ }\href {https://doi.org/10.1038/nphys4323} {\bibfield  {journal} {\bibinfo  {journal} {Nature Physics}\ }\textbf {\bibinfo {volume} {14}},\ \bibinfo {pages} {11} (\bibinfo {year} {2018})}\BibitemShut {NoStop}%
\bibitem [{\citenamefont {Ashida}\ \emph {et~al.}(2020)\citenamefont {Ashida}, \citenamefont {Gong},\ and\ \citenamefont {Ueda}}]{ashida2020nonhermitian}%
  \BibitemOpen
  \bibfield  {author} {\bibinfo {author} {\bibfnamefont {Y.}~\bibnamefont {Ashida}}, \bibinfo {author} {\bibfnamefont {Z.}~\bibnamefont {Gong}},\ and\ \bibinfo {author} {\bibfnamefont {M.}~\bibnamefont {Ueda}},\ }\href {https://doi.org/10.1080/00018732.2020.1716640} {\bibfield  {journal} {\bibinfo  {journal} {Advances in Physics}\ }\textbf {\bibinfo {volume} {69}},\ \bibinfo {pages} {249} (\bibinfo {year} {2020})}\BibitemShut {NoStop}%
\bibitem [{\citenamefont {Bender}(2007)}]{bender2007making}%
  \BibitemOpen
  \bibfield  {author} {\bibinfo {author} {\bibfnamefont {C.~M.}\ \bibnamefont {Bender}},\ }\href {https://doi.org/10.1088/0034-4885/70/6/R03} {\bibfield  {journal} {\bibinfo  {journal} {Reports on Progress in Physics}\ }\textbf {\bibinfo {volume} {70}},\ \bibinfo {pages} {947} (\bibinfo {year} {2007})}\BibitemShut {NoStop}%
\bibitem [{\citenamefont {Bennett}\ \emph {et~al.}(2002)\citenamefont {Bennett}, \citenamefont {Schatz}, \citenamefont {Rockwood},\ and\ \citenamefont {Wiesenfeld}}]{bennett2002huygens}%
  \BibitemOpen
  \bibfield  {author} {\bibinfo {author} {\bibfnamefont {M.}~\bibnamefont {Bennett}}, \bibinfo {author} {\bibfnamefont {M.}~\bibnamefont {Schatz}}, \bibinfo {author} {\bibfnamefont {H.}~\bibnamefont {Rockwood}},\ and\ \bibinfo {author} {\bibfnamefont {K.}~\bibnamefont {Wiesenfeld}},\ }\href {https://doi.org/10.1098/rspa.2001.0888} {\bibfield  {journal} {\bibinfo  {journal} {Philosophical Transactions of the Royal Society A}\ }\textbf {\bibinfo {volume} {458}},\ \bibinfo {pages} {563} (\bibinfo {year} {2002})}\BibitemShut {NoStop}%
\bibitem [{\citenamefont {Peña~Ramírez}\ \emph {et~al.}(2016)\citenamefont {Peña~Ramírez}, \citenamefont {Olvera}, \citenamefont {Nijmeijer},\ and\ \citenamefont {Álvarez}}]{penaramires2016sympathy}%
  \BibitemOpen
  \bibfield  {author} {\bibinfo {author} {\bibfnamefont {J.}~\bibnamefont {Peña~Ramírez}}, \bibinfo {author} {\bibfnamefont {L.~A.}\ \bibnamefont {Olvera}}, \bibinfo {author} {\bibfnamefont {H.}~\bibnamefont {Nijmeijer}},\ and\ \bibinfo {author} {\bibfnamefont {J.}~\bibnamefont {Álvarez}},\ }\href {https://doi.org/10.1038/srep23580} {\bibfield  {journal} {\bibinfo  {journal} {Scientific Reports}\ }\textbf {\bibinfo {volume} {6}},\ \bibinfo {pages} {23580} (\bibinfo {year} {2016})}\BibitemShut {NoStop}%
\bibitem [{\citenamefont {Engel}\ \emph {et~al.}(2007)\citenamefont {Engel}, \citenamefont {Calhoun}, \citenamefont {Read}, \citenamefont {Ahn}, \citenamefont {Mancal}, \citenamefont {Cheng}, \citenamefont {Blankenship},\ and\ \citenamefont {Fleming}}]{engel2007wavelike}%
  \BibitemOpen
  \bibfield  {author} {\bibinfo {author} {\bibfnamefont {G.~S.}\ \bibnamefont {Engel}}, \bibinfo {author} {\bibfnamefont {T.~R.}\ \bibnamefont {Calhoun}}, \bibinfo {author} {\bibfnamefont {E.~L.}\ \bibnamefont {Read}}, \bibinfo {author} {\bibfnamefont {T.-K.}\ \bibnamefont {Ahn}}, \bibinfo {author} {\bibfnamefont {T.}~\bibnamefont {Mancal}}, \bibinfo {author} {\bibfnamefont {Y.-C.}\ \bibnamefont {Cheng}}, \bibinfo {author} {\bibfnamefont {R.~E.}\ \bibnamefont {Blankenship}},\ and\ \bibinfo {author} {\bibfnamefont {G.~R.}\ \bibnamefont {Fleming}},\ }\href {https://doi.org/10.1038/nature05678} {\bibfield  {journal} {\bibinfo  {journal} {Nature}\ }\textbf {\bibinfo {volume} {446}},\ \bibinfo {pages} {782} (\bibinfo {year} {2007})}\BibitemShut {NoStop}%
\bibitem [{\citenamefont {Panitchayangkoon}\ \emph {et~al.}(2010)\citenamefont {Panitchayangkoon}, \citenamefont {Hayes}, \citenamefont {Fransted}, \citenamefont {Caram}, \citenamefont {Harel}, \citenamefont {Wen}, \citenamefont {Blankenship},\ and\ \citenamefont {Engel}}]{panitchayangkoon2010long}%
  \BibitemOpen
  \bibfield  {author} {\bibinfo {author} {\bibfnamefont {G.}~\bibnamefont {Panitchayangkoon}}, \bibinfo {author} {\bibfnamefont {D.}~\bibnamefont {Hayes}}, \bibinfo {author} {\bibfnamefont {K.~A.}\ \bibnamefont {Fransted}}, \bibinfo {author} {\bibfnamefont {J.~R.}\ \bibnamefont {Caram}}, \bibinfo {author} {\bibfnamefont {E.}~\bibnamefont {Harel}}, \bibinfo {author} {\bibfnamefont {J.~Z.}\ \bibnamefont {Wen}}, \bibinfo {author} {\bibfnamefont {R.~E.}\ \bibnamefont {Blankenship}},\ and\ \bibinfo {author} {\bibfnamefont {G.~S.}\ \bibnamefont {Engel}},\ }\href {https://doi.org/10.1073/pnas.1003062107} {\bibfield  {journal} {\bibinfo  {journal} {Proceedings of the National Academy of Sciences}\ }\textbf {\bibinfo {volume} {107}},\ \bibinfo {pages} {12766} (\bibinfo {year} {2010})}\BibitemShut {NoStop}%
\bibitem [{\citenamefont {Ishizaki}\ and\ \citenamefont {Fleming}(2009)}]{ishizaki2009theoretical}%
  \BibitemOpen
  \bibfield  {author} {\bibinfo {author} {\bibfnamefont {A.}~\bibnamefont {Ishizaki}}\ and\ \bibinfo {author} {\bibfnamefont {G.~R.}\ \bibnamefont {Fleming}},\ }\href {https://doi.org/10.1063/1.3149496} {\bibfield  {journal} {\bibinfo  {journal} {The Journal of Chemical Physics}\ }\textbf {\bibinfo {volume} {130}},\ \bibinfo {pages} {234111} (\bibinfo {year} {2009})}\BibitemShut {NoStop}%
\bibitem [{\citenamefont {Chin}\ \emph {et~al.}(2013)\citenamefont {Chin}, \citenamefont {Prior}, \citenamefont {Rosenbach}, \citenamefont {Caycedo-Soler}, \citenamefont {Huelga},\ and\ \citenamefont {Plenio}}]{chin2013role}%
  \BibitemOpen
  \bibfield  {author} {\bibinfo {author} {\bibfnamefont {A.~W.}\ \bibnamefont {Chin}}, \bibinfo {author} {\bibfnamefont {J.}~\bibnamefont {Prior}}, \bibinfo {author} {\bibfnamefont {R.}~\bibnamefont {Rosenbach}}, \bibinfo {author} {\bibfnamefont {F.}~\bibnamefont {Caycedo-Soler}}, \bibinfo {author} {\bibfnamefont {S.~F.}\ \bibnamefont {Huelga}},\ and\ \bibinfo {author} {\bibfnamefont {M.~B.}\ \bibnamefont {Plenio}},\ }\href {https://doi.org/10.1038/nphys2515} {\bibfield  {journal} {\bibinfo  {journal} {Nature Physics}\ }\textbf {\bibinfo {volume} {9}},\ \bibinfo {pages} {113} (\bibinfo {year} {2013})}\BibitemShut {NoStop}%
\bibitem [{\citenamefont {Tiwari}\ \emph {et~al.}(2013)\citenamefont {Tiwari}, \citenamefont {Peters},\ and\ \citenamefont {Jonas}}]{tiwari2013electronic}%
  \BibitemOpen
  \bibfield  {author} {\bibinfo {author} {\bibfnamefont {V.}~\bibnamefont {Tiwari}}, \bibinfo {author} {\bibfnamefont {W.~K.}\ \bibnamefont {Peters}},\ and\ \bibinfo {author} {\bibfnamefont {D.~M.}\ \bibnamefont {Jonas}},\ }\href {https://doi.org/10.1073/pnas.1217202110} {\bibfield  {journal} {\bibinfo  {journal} {Proceedings of the National Academy of Sciences}\ }\textbf {\bibinfo {volume} {110}},\ \bibinfo {pages} {1203} (\bibinfo {year} {2013})}\BibitemShut {NoStop}%
\bibitem [{\citenamefont {Hore}\ and\ \citenamefont {Mouritsen}(2016)}]{hore2016radical}%
  \BibitemOpen
  \bibfield  {author} {\bibinfo {author} {\bibfnamefont {P.~J.}\ \bibnamefont {Hore}}\ and\ \bibinfo {author} {\bibfnamefont {H.}~\bibnamefont {Mouritsen}},\ }\href {https://doi.org/10.1146/annurev-biophys-032116-094545} {\bibfield  {journal} {\bibinfo  {journal} {Annual Review of Biophysics}\ }\textbf {\bibinfo {volume} {45}},\ \bibinfo {pages} {299} (\bibinfo {year} {2016})}\BibitemShut {NoStop}%
\bibitem [{\citenamefont {Rodgers}\ and\ \citenamefont {Hore}(2009)}]{rodgers2009chemical}%
  \BibitemOpen
  \bibfield  {author} {\bibinfo {author} {\bibfnamefont {C.~T.}\ \bibnamefont {Rodgers}}\ and\ \bibinfo {author} {\bibfnamefont {P.~J.}\ \bibnamefont {Hore}},\ }\href {https://doi.org/10.1073/pnas.0711968106} {\bibfield  {journal} {\bibinfo  {journal} {Proc. Natl. Acad. Sci. USA}\ }\textbf {\bibinfo {volume} {106}},\ \bibinfo {pages} {353} (\bibinfo {year} {2009})}\BibitemShut {NoStop}%
\bibitem [{\citenamefont {Efimova}\ and\ \citenamefont {Hore}(2008)}]{efimova2008role}%
  \BibitemOpen
  \bibfield  {author} {\bibinfo {author} {\bibfnamefont {O.}~\bibnamefont {Efimova}}\ and\ \bibinfo {author} {\bibfnamefont {P.~J.}\ \bibnamefont {Hore}},\ }\href {https://doi.org/10.1529/biophysj.107.118564} {\bibfield  {journal} {\bibinfo  {journal} {Biophysical Journal}\ }\textbf {\bibinfo {volume} {94}},\ \bibinfo {pages} {1565} (\bibinfo {year} {2008})}\BibitemShut {NoStop}%
\bibitem [{\citenamefont {Player}\ and\ \citenamefont {Hore}(2019)}]{player2019viability}%
  \BibitemOpen
  \bibfield  {author} {\bibinfo {author} {\bibfnamefont {T.~C.}\ \bibnamefont {Player}}\ and\ \bibinfo {author} {\bibfnamefont {P.~J.}\ \bibnamefont {Hore}},\ }\href {https://doi.org/10.1063/1.5111570} {\bibfield  {journal} {\bibinfo  {journal} {Journal of Chemical Physics}\ }\textbf {\bibinfo {volume} {151}},\ \bibinfo {pages} {225101} (\bibinfo {year} {2019})}\BibitemShut {NoStop}%
\bibitem [{\citenamefont {Hiscock}\ and\ \citenamefont {Hore}(2016)}]{hiscock2016quantum}%
  \BibitemOpen
  \bibfield  {author} {\bibinfo {author} {\bibfnamefont {H.~G.}\ \bibnamefont {Hiscock}}\ and\ \bibinfo {author} {\bibfnamefont {P.~J.}\ \bibnamefont {Hore}},\ }\href {https://doi.org/10.1073/pnas.1600341113} {\bibfield  {journal} {\bibinfo  {journal} {Proceedings of the National Academy of Sciences USA}\ }\textbf {\bibinfo {volume} {113}},\ \bibinfo {pages} {4634} (\bibinfo {year} {2016})}\BibitemShut {NoStop}%
\bibitem [{\citenamefont {Anderson}(1954)}]{anderson1954stochastic}%
  \BibitemOpen
  \bibfield  {author} {\bibinfo {author} {\bibfnamefont {P.~W.}\ \bibnamefont {Anderson}},\ }\href {https://doi.org/10.1143/JPSJ.9.316} {\bibfield  {journal} {\bibinfo  {journal} {Journal of the Physical Society of Japan}\ }\textbf {\bibinfo {volume} {9}},\ \bibinfo {pages} {316} (\bibinfo {year} {1954})}\BibitemShut {NoStop}%
\bibitem [{\citenamefont {Kubo}(1954)}]{kubo1954resonance}%
  \BibitemOpen
  \bibfield  {author} {\bibinfo {author} {\bibfnamefont {R.}~\bibnamefont {Kubo}},\ }\href {https://doi.org/10.1143/JPSJ.9.935} {\bibfield  {journal} {\bibinfo  {journal} {Journal of the Physical Society of Japan}\ }\textbf {\bibinfo {volume} {9}},\ \bibinfo {pages} {935} (\bibinfo {year} {1954})}\BibitemShut {NoStop}%
\bibitem [{\citenamefont {Uhlenbeck}\ and\ \citenamefont {Ornstein}(1930)}]{uhlenbeck1930brownian}%
  \BibitemOpen
  \bibfield  {author} {\bibinfo {author} {\bibfnamefont {G.~E.}\ \bibnamefont {Uhlenbeck}}\ and\ \bibinfo {author} {\bibfnamefont {L.~S.}\ \bibnamefont {Ornstein}},\ }\href {https://doi.org/10.1103/PhysRev.36.823} {\bibfield  {journal} {\bibinfo  {journal} {Physical Review}\ }\textbf {\bibinfo {volume} {36}},\ \bibinfo {pages} {823} (\bibinfo {year} {1930})}\BibitemShut {NoStop}%
\bibitem [{\citenamefont {Tyagi}\ \emph {et~al.}(2024)\citenamefont {Tyagi}, \citenamefont {Li}, \citenamefont {Bittner}, \citenamefont {Piryatinski},\ and\ \citenamefont {Silva-Acu{\~n}a}}]{tyagi2024noise}%
  \BibitemOpen
  \bibfield  {author} {\bibinfo {author} {\bibfnamefont {B.}~\bibnamefont {Tyagi}}, \bibinfo {author} {\bibfnamefont {H.}~\bibnamefont {Li}}, \bibinfo {author} {\bibfnamefont {E.~R.}\ \bibnamefont {Bittner}}, \bibinfo {author} {\bibfnamefont {A.}~\bibnamefont {Piryatinski}},\ and\ \bibinfo {author} {\bibfnamefont {C.}~\bibnamefont {Silva-Acu{\~n}a}},\ }\href@noop {} {\bibfield  {journal} {\bibinfo  {journal} {The Journal of Physical Chemistry Letters}\ }\textbf {\bibinfo {volume} {15}},\ \bibinfo {pages} {10896} (\bibinfo {year} {2024})}\BibitemShut {NoStop}%
\bibitem [{\citenamefont {Bittner}\ \emph {et~al.}(2024)\citenamefont {Bittner}, \citenamefont {Li}, \citenamefont {Shah}, \citenamefont {Silva-Acu{\~n}a},\ and\ \citenamefont {Piryatinski}}]{bittner2024correlated}%
  \BibitemOpen
  \bibfield  {author} {\bibinfo {author} {\bibfnamefont {E.~R.}\ \bibnamefont {Bittner}}, \bibinfo {author} {\bibfnamefont {H.}~\bibnamefont {Li}}, \bibinfo {author} {\bibfnamefont {S.~A.}\ \bibnamefont {Shah}}, \bibinfo {author} {\bibfnamefont {C.}~\bibnamefont {Silva-Acu{\~n}a}},\ and\ \bibinfo {author} {\bibfnamefont {A.}~\bibnamefont {Piryatinski}},\ }\href@noop {} {\bibfield  {journal} {\bibinfo  {journal} {Philosophical Magazine}\ }\textbf {\bibinfo {volume} {104}},\ \bibinfo {pages} {630} (\bibinfo {year} {2024})}\BibitemShut {NoStop}%
\bibitem [{\citenamefont {Bittner}\ and\ \citenamefont {Tyagi}(2025{\natexlab{a}})}]{bittner2025noise}%
  \BibitemOpen
  \bibfield  {author} {\bibinfo {author} {\bibfnamefont {E.~R.}\ \bibnamefont {Bittner}}\ and\ \bibinfo {author} {\bibfnamefont {B.}~\bibnamefont {Tyagi}},\ }\href@noop {} {\bibfield  {journal} {\bibinfo  {journal} {The Journal of Chemical Physics}\ }\textbf {\bibinfo {volume} {162}} (\bibinfo {year} {2025}{\natexlab{a}})}\BibitemShut {NoStop}%
\bibitem [{\citenamefont {Bittner}\ and\ \citenamefont {Tyagi}(2025{\natexlab{b}})}]{bittner2025statistical}%
  \BibitemOpen
  \bibfield  {author} {\bibinfo {author} {\bibfnamefont {E.~R.}\ \bibnamefont {Bittner}}\ and\ \bibinfo {author} {\bibfnamefont {B.}~\bibnamefont {Tyagi}},\ }\href@noop {} {\bibfield  {journal} {\bibinfo  {journal} {arXiv preprint arXiv:2504.02173}\ } (\bibinfo {year} {2025}{\natexlab{b}})}\BibitemShut {NoStop}%
\bibitem [{\citenamefont {Ott}\ and\ \citenamefont {Antonsen}(2008)}]{ott2008low}%
  \BibitemOpen
  \bibfield  {author} {\bibinfo {author} {\bibfnamefont {E.}~\bibnamefont {Ott}}\ and\ \bibinfo {author} {\bibfnamefont {T.~M.}\ \bibnamefont {Antonsen}},\ }\href@noop {} {\bibfield  {journal} {\bibinfo  {journal} {Chaos: An Interdisciplinary Journal of Nonlinear Science}\ }\textbf {\bibinfo {volume} {18}},\ \bibinfo {pages} {037113} (\bibinfo {year} {2008})}\BibitemShut {NoStop}%
\bibitem [{\citenamefont {Lee}\ \emph {et~al.}(2011)\citenamefont {Lee}, \citenamefont {Cross},\ and\ \citenamefont {Sadeghpour}}]{lee2011quantum}%
  \BibitemOpen
  \bibfield  {author} {\bibinfo {author} {\bibfnamefont {T.~E.}\ \bibnamefont {Lee}}, \bibinfo {author} {\bibfnamefont {M.~C.}\ \bibnamefont {Cross}},\ and\ \bibinfo {author} {\bibfnamefont {H.~R.}\ \bibnamefont {Sadeghpour}},\ }\href@noop {} {\bibfield  {journal} {\bibinfo  {journal} {Physical Review Letters}\ }\textbf {\bibinfo {volume} {106}},\ \bibinfo {pages} {143001} (\bibinfo {year} {2011})}\BibitemShut {NoStop}%
\bibitem [{\citenamefont {Teramae}\ and\ \citenamefont {Tanaka}(2004)}]{teramae2004robustness}%
  \BibitemOpen
  \bibfield  {author} {\bibinfo {author} {\bibfnamefont {J.-n.}\ \bibnamefont {Teramae}}\ and\ \bibinfo {author} {\bibfnamefont {D.}~\bibnamefont {Tanaka}},\ }\href@noop {} {\bibfield  {journal} {\bibinfo  {journal} {Physical Review Letters}\ }\textbf {\bibinfo {volume} {93}},\ \bibinfo {pages} {204103} (\bibinfo {year} {2004})}\BibitemShut {NoStop}%
\bibitem [{\citenamefont {Nakao}\ \emph {et~al.}(2007)\citenamefont {Nakao}, \citenamefont {Arai}, \citenamefont {Kawamura},\ and\ \citenamefont {Kuramoto}}]{nakao2007noise}%
  \BibitemOpen
  \bibfield  {author} {\bibinfo {author} {\bibfnamefont {H.}~\bibnamefont {Nakao}}, \bibinfo {author} {\bibfnamefont {K.}~\bibnamefont {Arai}}, \bibinfo {author} {\bibfnamefont {Y.}~\bibnamefont {Kawamura}},\ and\ \bibinfo {author} {\bibfnamefont {Y.}~\bibnamefont {Kuramoto}},\ }\href@noop {} {\bibfield  {journal} {\bibinfo  {journal} {Physical Review Letters}\ }\textbf {\bibinfo {volume} {98}},\ \bibinfo {pages} {184101} (\bibinfo {year} {2007})}\BibitemShut {NoStop}%
\bibitem [{\citenamefont {Braun}(2002)}]{braun2002creation}%
  \BibitemOpen
  \bibfield  {author} {\bibinfo {author} {\bibfnamefont {D.}~\bibnamefont {Braun}},\ }\href@noop {} {\bibfield  {journal} {\bibinfo  {journal} {Physical Review Letters}\ }\textbf {\bibinfo {volume} {89}},\ \bibinfo {pages} {277901} (\bibinfo {year} {2002})}\BibitemShut {NoStop}%
\bibitem [{\citenamefont {Benatti}\ \emph {et~al.}(2003)\citenamefont {Benatti}, \citenamefont {Floreanini},\ and\ \citenamefont {Piani}}]{benatti2006entangling}%
  \BibitemOpen
  \bibfield  {author} {\bibinfo {author} {\bibfnamefont {F.}~\bibnamefont {Benatti}}, \bibinfo {author} {\bibfnamefont {R.}~\bibnamefont {Floreanini}},\ and\ \bibinfo {author} {\bibfnamefont {M.}~\bibnamefont {Piani}},\ }\href@noop {} {\bibfield  {journal} {\bibinfo  {journal} {Physical Review Letters}\ }\textbf {\bibinfo {volume} {91}},\ \bibinfo {pages} {070402} (\bibinfo {year} {2003})}\BibitemShut {NoStop}%
\bibitem [{\citenamefont {Benedetti}\ \emph {et~al.}(2012)\citenamefont {Benedetti}, \citenamefont {Buscemi}, \citenamefont {Bordone},\ and\ \citenamefont {Paris}}]{benedetti2012effects}%
  \BibitemOpen
  \bibfield  {author} {\bibinfo {author} {\bibfnamefont {C.}~\bibnamefont {Benedetti}}, \bibinfo {author} {\bibfnamefont {F.}~\bibnamefont {Buscemi}}, \bibinfo {author} {\bibfnamefont {P.}~\bibnamefont {Bordone}},\ and\ \bibinfo {author} {\bibfnamefont {M.~G.~A.}\ \bibnamefont {Paris}},\ }\href@noop {} {\bibfield  {journal} {\bibinfo  {journal} {International Journal of Quantum Information}\ }\textbf {\bibinfo {volume} {10}},\ \bibinfo {pages} {1241005} (\bibinfo {year} {2012})}\BibitemShut {NoStop}%
\end{thebibliography}%

\end{document}